\documentclass[prl,aps,twocolumn]{revtex4-2}

\usepackage{bm,bbm,color,float,dcolumn,amsmath,amssymb,graphicx,subfigure}

\usepackage[colorlinks=true]{hyperref}
\hypersetup{linkcolor=blue,citecolor=blue,urlcolor=blue}

\definecolor{psychedelicpurple}{rgb}{0.87, 0.0, 1.0}
\definecolor{violet}{rgb}{0.56, 0, 1}

\begin{document}

\title{Extracting the Luttinger parameter from a single wave function}

\author{Bi-Yang Tan$^{1}$}
\thanks{These two authors contributed equally.}

\author{Yueshui Zhang$^{2}$}
\thanks{These two authors contributed equally.}

\author{Hua-Chen Zhang$^{3}$, Wei Tang$^{4}$, \\ Lei Wang$^{5,6}$}

\author{Hong-Hao Tu$^{2}$}
\email{h.tu@lmu.de}

\author{Ying-Hai Wu$^{1}$}
\email{yinghaiwu88@hust.edu.cn}

\affiliation
{
$^1$ School of Physics and Wuhan National High Magnetic Field Center, Huazhong University of Science and Technology, Wuhan 430074, China \\
$^2$ Physics Department and Arnold Sommerfeld Center for Theoretical Physics, Ludwig-Maximilians-Universit\"at M\"unchen, 80333 Munich, Germany \\
$^3$ Department of Physics and Astronomy, Aarhus University, DK-8000 Aarhus C, Denmark \\
$^4$ Department of Physics and Astronomy, Ghent University, Krijgslaan 281, S9, B-9000 Ghent, Belgium \\
$^5$ Institute of Physics, Chinese Academy of Sciences, Beijing 100190, China \\
$^6$ Songshan Lake Materials Laboratory, Dongguan, Guangdong 523808, China
}

\date{\today}

\begin{abstract}
The low-energy physics of Tomonaga-Luttinger liquids (TLLs) is controlled by the Luttinger parameter. We demonstrate that this parameter can be extracted from a single wave function for one-component TLLs with periodic boundary condition. This method relies on the fact that TLLs are described by conformal field theory in which crosscap states can be constructed. The overlaps between the crosscap states and the ground state as well as some excited states are proved to be universal numbers that directly reveal the Luttinger parameter. In microscopic lattice models, crosscap states are formed by putting each pair of antipodal sites into a maximally entangled state. Analytical and numerical calculations are performed in a few representative models to substantiate the conformal field theory prediction. The extracted Luttinger parameters are generally quite accurate in finite-size systems with moderate lengths, so there is no need to perform data fitting and/or finite-size scaling.
\end{abstract}

\maketitle

{\em Introduction} --- The theory of Tomonaga-Luttinger liquid (TLL) is a great triumph of strongly correlated physics~\cite{Giamarchi-Book}. Based on previous work of Tomonaga~\cite{Tomonaga1950}, Luttinger introduced a model to study interacting fermions in one dimension~\cite{Luttinger1963}. It clearly exemplifies the peculiarity of reduced dimensionality because infinitesimal interaction in a one-dimensional system drives it away from the Fermi liquid. In subsequent works~\cite{Mattis1965,Mattis1974,Luther1974,Haldane1981,vonDelft1998}, the bosonization framework was established to provide a unified treatment for many strongly correlated problems in one dimension. Spin chains are usually studied using the Jordan-Wigner transformation that converts spins to fermions. Gapless phases of bosons or fermions can be understood using free boson fields that correspond to density fluctuations. Instabilities of the TLL lead to symmetry-breaking ordered phases. It is not easy to realize one-dimensional systems, but experimental investigations have been carried out in solid state and cold atom platforms~\cite{Yacoby1996,Schwartz1998,Bockrath1999,YaoZ1999,Levy2006,Kono2015,Hoffer2008,Haller2010,YangB2017,Vijayan2020}. Another context in which TLL thrives is the edge of certain two-dimensional topological states~\cite{WenXG1995,ChangAM2003,WuCJ2006,LiTX2015}.

The low-energy physics of one-component TLLs is captured by the Hamiltonian
\begin{align}
H = \frac{v}{8\pi} \int^{L}_{0} \mathrm{d}x \, \left[ (\partial_{x}\hat{\varphi})^{2} + (\partial_{x}\hat{\theta})^{2} \right],
\label{eq:LuttHami}
\end{align}
which is a system of compactified bosons in the language of conformal field theory (CFT). Here $L$ is the length of the system, $v$ is the characteristic velocity, $\hat{\varphi}$ is a compactified boson field with radius $R$ (i.e., $\hat{\varphi} \sim \hat{\varphi} + 2\pi R$), and $\hat{\theta}$ is the conjugate boson field of $\hat{\varphi}$ with radius $R^{\prime}=2/R$. The Luttinger parameter $K$ can be expressed using the radius as $K=R^{2}/4$. Its value is $1$ for free fermions and deviates from $1$ when interaction is turned on. For a given TLL, many properties are determined by $K$ so it is routinely computed in numerical studies. If the Hamiltonian has a microscopic U(1) symmetry (e.g., particle number conservation), $K$ can be obtained by comparing the ground-state energy of multiple systems with different U(1) charges~\cite{Alcaraz1987}. This is feasible because $K$ can be interpreted as the stiffness of a superfluid. The powerlaw decay of certain correlation functions can be used to deduce $K$ when their bosonized expressions are known~\cite{Giamarchi-Book}. It has been found that $K$ also appears in the U(1) charge fluctuations of a subsystem~\cite{SongHF2010,Nishimoto2011}, the U(1) symmetry-resolved entanglement spectrum~\cite{laeuchli2013,Vanders2019}, different types of entanglement entropy~\cite{Calabrese2009,Herdman2015,Barghathi2017,Bastianello2019,Estienne2024}, and finite entanglement scaling analysis~\cite{HeYC2021,Eberharter2023}.

In this Letter, we propose that the Luttinger parameter in microscopic models can be extracted from the eigenstate crosscap overlap (ESCO). It has three favorable features: i) a single wave function instead of multiple states is used; ii) data fitting or finite-size scaling is not needed; iii) microscopic U(1) symmetry is not required. To the best of our knowledge, no other method enjoys all three advantages simultaneously. Our method is motivated by classical results and recent progresses in CFT and integrable models. When a CFT is defined on the Klein bottle, important information can be inferred from the thermal entropy~\cite{TuHH2017,TangW2017,ChenL2017,WangHX2018,TangW2019,LiZQ2020,tangW2020,Garcia2021,ChenY2020,LiH2021,Vanhove2022,ZhangYS2023,Shimizu2024}. The Klein bottle can be viewed as a special cylinder with crosscaps at its ends. A simple circle with all pairs of antipodal points identified becomes a crosscap. It is possible to define a variety of crosscap states in the framework of boundary CFT~\cite{Ishibashi1989}. For a rational CFT, the crosscap states have universal overlaps (``crosscap coefficients'') with its ground state~\cite{Fioravanti1994,Pradisi1996,Fuchs2000}. In some integrable models, the crosscap states have been identified as ``integrable boundary states''~\cite{Caetano2022,Ekman2022,Gombor2022,Gombor2023,HeM2023}. The overlaps between all eigenstates of these integrable models and the crosscap states can be expressed as determinants, but their physical significance is not transparent. We aim to establish a general framework for studying crosscap states in CFT (without and with perturbations) such that physical properties can be deduced from their overlaps with eigenstates.

\begin{figure}[ht]
\centering
\includegraphics[width=0.48\textwidth]{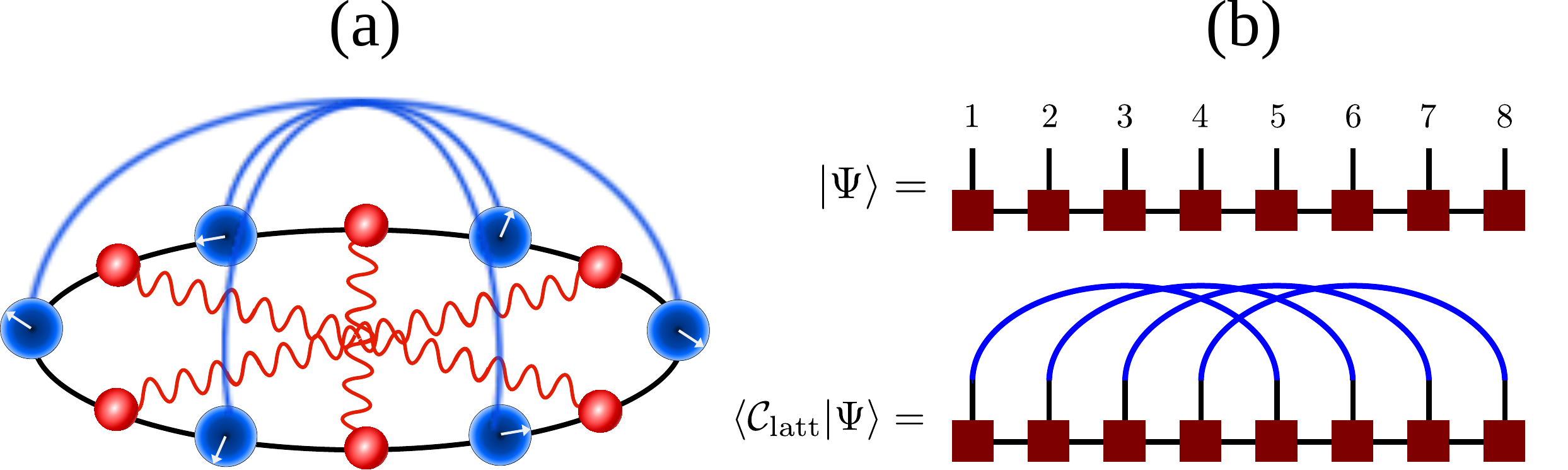}
\caption{(a) Schematics of the modified Villain model and its crosscap states $|\mathcal{C}_{\pm}\rangle$. (b) Graphic representations of an MPS with eight sites and its crosscap overlap.}
\label{fig:Figure1}
\end{figure}

{\em General construction of crosscap states} --- While the crosscap is easy to visualize, the problem of what kind of states qualify as conformal crosscap states in CFTs has not been completely settled~\cite{Fioravanti1994,Pradisi1996,Fuchs2000}. We take the viewpoint that a conformal crosscap state should satisfy two sets of constraints. First, the field operators on antipodal spatial coordinates are identified with each other in some way. If a CFT has Lagrangian description, this can be made precise using the path integral formalism~\cite{Blumenhagen-Book,Brunner2004,ChoGY2015}. These requisites are physical manifestations of the geometric structure of crosscap and referred to as ``sewing conditions". Second, the partition functions on a cylinder with crosscap boundaries (tree channel) and on a Klein bottle (loop channel) should be equal. This ``loop-tree channel correspondence'' allows one to determine certain coefficients in crosscap states. As we turn to lattice models, the problem becomes even more complicated because the lattice counterpart of a field operator is generally difficult to find. It is thus unlikely that a conformal crosscap state can be converted rigorously to a lattice one. Based on the intuitive picture of identifying antipodal spatial points, recent works have studied lattice crosscap states in which antipodal sites form Bell states~\cite{Caetano2022,Ekman2022,Gombor2022,Gombor2023,HeM2023}, but their relation to the conformal crosscap states has not been elucidated. 

It is particularly instructive to study a specific lattice realization of the Hamiltonian in Eq.~\eqref{eq:LuttHami}, namely the modified Villain model ~\cite{ChengM2023} as illustrated in Fig.~\ref{fig:Figure1}(a)
\begin{align}
    H_{\mathrm{mV}} = \sum^{N}_{j=1} \left[ (2\pi R\,\hat{p}_{j})^{2} + {R^{\prime}}^{2} (\hat{X}_{j+1}-\hat{X}_{j}+2\pi \hat{n}_{j,j+1})^2 \right]. 
\label{eq:mVHami-1}
\end{align}
There are has $N$ sites in the chain (labeled by $j$), the lattice spacing is $a$, and periodic boundary condition is imposed. $\hat{X}_{j}$ ($\hat{E}_{j,j+1}$) is the site (link) variable at site $j$ [link $(j,j+1)$] and $\hat{p}_{j}$ ($\hat{n}_{j,j+1}$) is the associated canonical momentum, and they satisfy the standard commutation relations $[\hat{X}_{j},\hat{p}_{l}]=[\hat{E}_{j,j+1},\hat{n}_{l,l+1}]=i\delta_{jl}$. The eigenbasis of these variables are denoted as $|X_{j}\rangle$ ($X_{j} \in \mathbb{R}$) and $|E_{j,j+1}\rangle$ [$E_{j,j+1}\in [0,2\pi)$], which span the local Hilbert space at site $j$ and link $(j,j+1)$, respectively. The tensor product of these local ones is the full Hilbert space. However, there are additional local operators $\hat{G}_j=e^{i (\hat{E}_{j,j+1}-\hat{E}_{j-1,j}-2\pi \hat{p}_j)}$ that commute with the Hamiltonian. We define the physical subspace $\mathcal{H}_{\mathrm{phy}}$ by imposing the Gauss law constraints $\hat{G}_{j}=1 \; \forall j$. The projection operator onto the subspace $\mathcal{H}_{\mathrm{phy}}$ is denoted as $\mathcal{P}_{G}$. Using new variables $\hat{\theta}_j =R^{\prime}(\hat{X}_j+2\pi \sum_{k=1}^j \hat{n}_{k-1,k})$ and $\hat{\varphi}_{j,j+1}=R \hat{E}_{j,j+1}$~\cite{ChengM2023}, Eq.~\eqref{eq:mVHami-1} can be rewritten as
\begin{align}
H_{\mathrm{mV}} =\sum^{N}_{j=1} \left[ (\hat{\varphi}_{j+1,j+2}-\hat{\varphi}_{j,j+1})^{2} +(\hat{\theta}_{j+1}-\hat{\theta}_{j})^{2} \right] \, . 
\label{eq:mVHami-2}
\end{align}
Its continuum limit is taken by sending $N\rightarrow \infty$ and $a\rightarrow 0$ with $L=Na$ kept unchanged. After the substitutions $\hat{\varphi}_{j+1,j+2}-\hat{\varphi}_{j,j+1} \rightarrow a\partial_{x}\hat{\varphi} (x)$ and $\hat{\theta}_{j+1}-\hat{\theta}_{j} \rightarrow a\partial_x \hat{\theta} (x)$, the TLL Hamiltonian emerges from Eq.~\eqref{eq:mVHami-2}. The operators $e^{i\hat{\theta}_j/R^{\prime}}$ and $e^{i\hat{\varphi}_{j,j+1}/R}$ act within $\mathcal{H}_{\mathrm{phy}}$ thanks to the commutation relations $[e^{i\hat{\theta}_j/R^{\prime}},\hat{G}_{j^{\prime}}]=[e^{i\hat{\varphi}_{j,j+1}/R},\hat{G}_{j^{\prime}}]=0 \; \forall j,j^{\prime}$. The compactification radii of $\hat{\theta}_j$ and $\hat{\varphi}_{j,j+1}$ are $R^{\prime}$ and $R$, respectively. 

Motivated by the intuitive picture mentioned above and previous works~\cite{Fioravanti1994,Pradisi1996,Fuchs2000,Caetano2022,Ekman2022,Gombor2022,Gombor2023,HeM2023}, we construct the Ans\"atze
\begin{align}
    |\mathcal{C}_{\pm}\rangle = \mathcal{P}_{G}  \prod^{N/2}_{j=1} |\mathcal{B}^{\mathrm{V}}_{\pm}\rangle_{j,j+N/2} \otimes |\mathcal{B}^{\mathrm{L}}_{\mp}\rangle_{j,j+N/2} \,
\label{eq:crosscap-1}
\end{align}
for crosscap states by putting antipodal links and sites in the generalized Bell pairs
\begin{align}
|\mathcal{B}^{\mathrm{L}}_{\mp}\rangle_{j,j+N/2} &= \int^{2\pi}_{0} \mathrm{d}E \, \vert E\rangle_{j,j+1} |\mp E\rangle_{j+N/2,j+N/2+1} \, , \nonumber\\
|\mathcal{B}^{\mathrm{V}}_{\pm}\rangle_{j,j+N/2} &= \int^{\infty}_{-\infty} \mathrm{d}X \, \vert X\rangle_{j} |\pm X\rangle_{j+N/2} \, ,
\label{eq:Bell-1}
\end{align}
respectively. It can be verified that the relations 
\begin{align}
    e^{i(\hat{\theta}_j \mp \hat{\theta}_{j+N/2})/R^{\prime}}|\mathcal{C}_{\pm}\rangle &= |\mathcal{C}_\pm\rangle \, ,  \nonumber \\
    e^{i(\hat{\varphi}_{j,j+1} \pm \hat{\varphi}_{j+N/2,j+N/2+1})/R}|\mathcal{C}_{\pm}\rangle &= |\mathcal{C}_\pm\rangle \,
\label{eq:gluing1}
\end{align}
are fulfilled for all $j$, which in the continuum limit becomes
\begin{align}
    e^{i[\hat{\theta}(x) \mp \hat{\theta}(x+L/2)]/R^{\prime}} |\mathcal{C}_{\pm}\rangle &= |\mathcal{C}_\pm\rangle \, ,  \nonumber \\
    e^{i[\hat{\varphi}(x) \pm \hat{\varphi}(x+L/2)]/R}|\mathcal{C}_{\pm}\rangle &= |\mathcal{C}_\pm\rangle \,.
\label{eq:gluing2}
\end{align}
These identities are sewing conditions in the sense that field operators on antipodal spatial coordinates are combined to form certain operators under which the crosscap states are invariant. It is amusing that they can be derived rigorously using our formalism.

For subsequent calculations, we introduce the mode expansions~\cite{Francesco-Book}
\begin{align}
    \hat{\theta}(x) &= \tilde{\varphi}_{0} + \frac{4\pi x}{L}\pi_{0} + \sum_{k\neq 0}\frac{i}{k} \left( e^{i\frac{2\pi k}{L}x} a_{k} - e^{-i\frac{2\pi k}{L}x}\bar{a}_{k} \right), \nonumber \\
    \hat{\varphi}(x) &= \varphi_{0} + \frac{4\pi x}{L}\tilde{\pi}_{0} + \sum_{k\neq 0}\frac{i}{k} \left( e^{i\frac{2\pi k}{L}x}a_{k} + e^{-i\frac{2\pi k}{L}x}\bar{a}_{k} \right),
\label{eq:mode-expan}
\end{align}
where the zero modes and oscillatory modes satisfy $[\varphi_{0},\pi_{0}]=[\tilde{\varphi}_{0},\tilde{\pi}_{0}]=i$ and $[a_{k},a_{l}]=[\bar{a}_{k},\bar{a}_{l}]=k\delta_{k+l,0}$, respectively. $\varphi_{0}$ ($\tilde{\varphi}_{0}$) is an angular variable with radius $R$ ($R^{\prime}$), and $\pi_{0}$ ($\tilde{\pi}_{0}$) is the associated canonical momentum. The eigenvectors of $\pi_{0}$ and $\tilde{\pi}_{0}$ that can be annihilated by all bosonic annihilation operators $a_{k>0}$ and $\bar{a}_{k>0}$ are denoted as $|n,m\rangle$ ($n,m\in\mathbb{Z}$); they are highest weight states (Virasoro primary states) of the CFT and satisfy $\pi_{0} |n,m\rangle = \tfrac{n}{R}|n,m\rangle$ and $\tilde{\pi}_{0} |n,m\rangle = \tfrac{m}{R^{\prime}}|n,m\rangle$. The TLL Hamiltonian in Eq.~\eqref{eq:LuttHami} becomes 
\begin{align}
    H =\frac{2\pi v}{L} \left[\pi^{2}_{0} + \tilde{\pi}^{2}_{0} + \sum^{\infty}_{k=1} (a_{-k}a_{k} + \bar{a}_{-k}\bar{a}_{k}) - \frac{1}{12} \right],
\end{align}
whose energy eigenstates are just Virasoro primary states $|n,m\rangle$ and the descendant states obtained by applying the creation operators $a_{k<0}$ and $\bar{a}_{k<0}$ on top of $|n,m\rangle$. The crosscap states can be expressed as
\begin{align}
    |\mathcal{C}_{+}\rangle &= \kappa_{+} \exp\left[\sum^{\infty}_{k=1} \frac{(-1)^{k}}{k} a_{-k}\bar{a}_{-k} \right] \sum_{n\in\mathbb{Z}}|2n,0\rangle \, ,\nonumber \\
    |\mathcal{C}_{-}\rangle &= \kappa_{-} \exp\left[-\sum^{\infty}_{k=1} \frac{(-1)^{k}}{k} a_{-k}\bar{a}_{-k} \right] \sum_{m\in\mathbb{Z}}|0,2m\rangle.
\label{eq:cross-field}
\end{align}
The overall factors are not fixed by normalization but should be chosen as $\kappa_+ =\sqrt{R^{\prime}}$ and $\kappa_- =\sqrt{R}$ to fulfill the loop-tree channel correspondence~\cite{Append}. 

The conformal crosscap states in Eq.~\eqref{eq:cross-field} are scale-invariant (i.e., they contain no scale) and cannot be normalized. Their lattice counterparts may be normalizable [Eq.~\eqref{eq:cross-lattice} for spin-1/2 chains defined below] or not normalizable [Eq.~\eqref{eq:crosscap-1} for the modified Villain model] , depending on the local Hilbert space dimension. In fact, for a given lattice model whose low-energy theory is the compactified boson CFT, the crosscap states together with the coefficients $\kappa_\pm$ in Eq.~\eqref{eq:cross-field} should be understood as the continuum description of their lattice counterparts $|\mathcal{C}_{\mathrm{latt}}^\pm\rangle$. This means that they are fully determined by the overlaps of lattice crosscap states with low-lying eigenstates in the continuum limit: $\langle \mathcal{C}_\pm |n,m; \{n_k\},\{\bar{n}_k\}\rangle = \lim_{N\to\infty} \langle\mathcal{C}_{\mathrm{latt}}^\pm |\psi_{n,m; \{n_k\},\{\bar{n}_k\}}\rangle$, where $|\psi_{n,m; \{n_k\},\{\bar{n}_k\}}\rangle$ is the lattice counterpart of the CFT eigenstate $|n,m; \{n_k\},\{\bar{n}_k\}\rangle$.

An appealing property of $|\mathcal{C}_{\pm}\rangle$ is that their overlaps with certain highest weight states $|n,m\rangle$ are universal numbers depending only on the compactification radius (equivalently, the Luttinger parameter):
\begin{align}
    |\langle\mathcal{C_+}|2n,0\rangle|^{2} = R^{\prime} = \frac{1}{\sqrt{K}},\quad n\in\mathbb{Z} \, , \nonumber \\
    |\langle\mathcal{C_-}|0,2m\rangle|^{2} = R = 2\sqrt{K},\quad m\in\mathbb{Z} \, .
\label{eq:cross-overlap}
\end{align}
It was found in Ref.~\cite{TangW2019} that the Luttinger parameter appears in the thermal entropy of compactified boson CFTs on the Klein bottle. In view of the spacetime symmetry of CFTs, it is natural to expect that the same variable can also be found using the ground state alone. Our derivation not only puts this expectation on a firm ground but also provides further insights into crosscap states. Some excited states are incorporated in the present formalism, and perturbations to a pure CFT can be investigated~\cite{ZhangYS2024}.

{\em Spin chains with microscopic U(1) symmetry} --- To apply the general theory in specific lattice models, the crosscap states should be expressed in terms of the microscopic degrees of freedom. It is helpful to begin with spin-1/2 chains. The first one is the XXZ model
\begin{align}
H_{\mathrm{XXZ}} = \sum^{N}_{j=1} \left( S^{x}_{j} S^{x}_{j+1} + S^{y}_{j} S^{y}_{j+1} + \Delta S^{z}_{j} S^{z}_{j+1} \right)
\label{eq:XXZ-chain}
\end{align}
with $K=\frac{\pi}{2(\pi-\cos^{-1}\Delta)}$ for $\Delta \in (-1,1]$~\cite{Giamarchi-Book}. The second one is the Haldane-Shastry (HS) model
\begin{align}
H_{\mathrm{HS}} = \left(\frac{\pi}{N}\right)^{2} \sum_{1\leq j<l \leq N} \frac{\mathbf{S}_{j} \cdot \mathbf{S}_{l}}{ \sin^{2} \left[\frac{\pi}{N}(j-l)\right]}
\label{eq:HS-chain}
\end{align}
with $K=1/2$~\cite{Haldane1988a,Shastry1988}. Both models have a U(1) symmetry associated with the conservation of the $z$-component of the total spin.

It is natural to speculate that
\begin{align}
|\mathcal{C}_{\mathrm{latt}}\rangle = \prod^{N/2}_{j=1} \left(|\uparrow\rangle_{j} |\uparrow\rangle_{j+N/2} + |\downarrow\rangle_j |\downarrow\rangle_{j+N/2} \right)
\label{eq:cross-lattice}
\end{align}
is a crosscap state when $N$ is even~\cite{Caetano2022,Ekman2022,Gombor2022,Gombor2023,HeM2023}, but there is one subtlety: it corresponds to $|\mathcal{C}_{+}\rangle$ only if $\mathrm{mod}(N,4)=0$ but does not for the cases with $\mathrm{mod}(N,4)=2$. This property can be traced back to the constraint $(S^{z}_{j}-S^{z}_{j+N/2})|\mathcal{C}_{\mathrm{latt}}\rangle=0$. The $S^{z}_{j}$ operator in bosonization reads $S^{z}_{j}\sim\partial_{x}\hat{\theta}(x)+(-1)^{j}\alpha\cos[\hat{\theta}(x)/R^{\prime}]$ for XXZ and HS chains ($\alpha$ is a model-dependent constant). For $\mathrm{mod}(N,4)=0$, the lattice constraint $(S^{z}_{j}-S^{z}_{j+N/2})|\mathcal{C}_{\mathrm{latt}}\rangle=0$ is hence consistent with Eq.~\eqref{eq:gluing2}.

The validity of Eq.~\eqref{eq:cross-lattice} in the XY model (i.e., $\Delta=0$ in the XXZ model) and the HS model can be substantiated by analytical calculations. In the hardcore boson basis, the Virasoro primary states $|n,0\rangle$ have the wave functions $\Psi^\lambda (x_{1},\ldots,x_{M}) \propto (-1)^{\sum_{l} x_{l}} \prod_{j<k} \left| \sin\frac{\pi(x_{j}-x_{k})}{N} \right|^{\lambda}$, where $1\leq x_{1} <\cdots< x_{M} \leq N$ denote the positions of spin-$\uparrow$ sites (occupied by hardcore bosons), and $\lambda=1$ ($\lambda=2$) for the XY (HS) chain~\cite{Kuramoto-Book}. The number of bosons $M$ is related to the U(1) charge $n$ via $M=N/2+n$. It is important to note that the cases with $\mathrm{mod}(N,4)=2$ and $\mathrm{mod}(N,4)=0$ are different. Because $|\mathcal{C}_{\mathrm{latt}}\rangle$ only contains states with even $M$, $\langle\mathcal{C}_{\mathrm{latt}} \vert \Psi^{\lambda}(M)\rangle$ vanishes identically for odd $M$. We also have $\langle \mathcal{C}_{\mathrm{latt}} \vert \Psi^{\lambda}(M=N/2+2n)\rangle =0$ when $\mathrm{mod}(N,4)=2$, so $|\mathcal{C}_{\mathrm{latt}}\rangle$ does not correspond to $|\mathcal{C}_{\pm}\rangle$ in these cases. In contrast, it can be proven that 
\begin{align}
| \langle\mathcal{C}_{\mathrm{latt}} | \Psi^{\lambda=1}(M) \rangle|^{2} &= 1 \, , \nonumber \\ 
| \langle\mathcal{C}_{\mathrm{latt}} | \Psi^{\lambda=2}(M) \rangle|^{2} &= \frac{2^{M} (M!)^{3}}{[(M/2)!]^{2} (2M)!} = \sqrt{2}+\mathcal{O}(1/M) \, 
\end{align}
if $\mathrm{mod}(N,4)=0$ and $M=N/2+2n$~\cite{Append}. This is a perfect agreement with the field theory prediction Eq.~\eqref{eq:cross-overlap}, giving a strong indication that $|\mathcal{C}_{\mathrm{latt}}\rangle$ corresponds to $|\mathcal{C}_+\rangle$ for $\mathrm{mod}(N,4)=0$.

\begin{figure}[ht]
\centering
\includegraphics[width=0.48\textwidth]{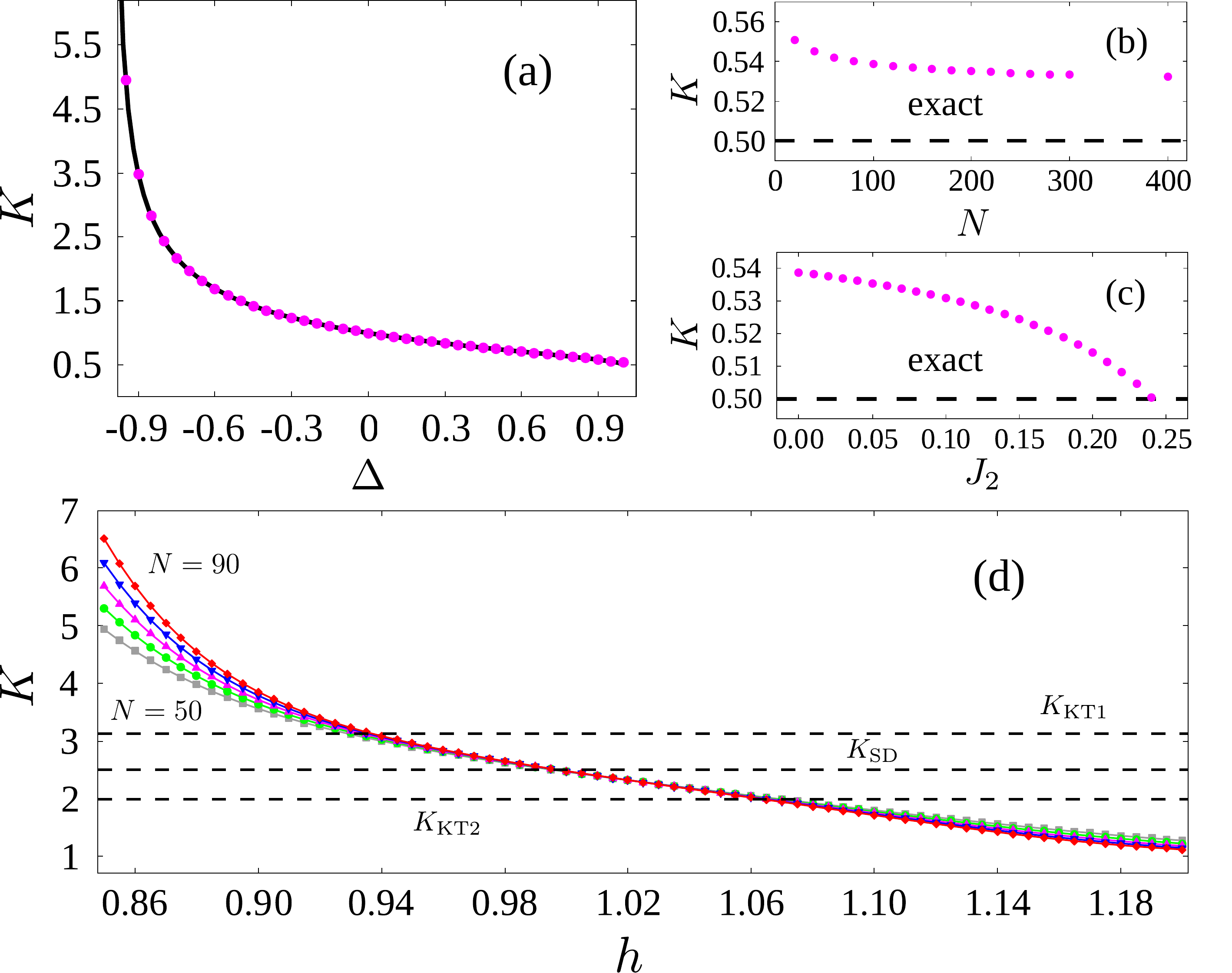}
\caption{Numerically extracted Luttinger parameters in three models. Exact values are indicated using solid or dashed black lines for comparison. (a) The XXZ model with $N=100$ and various $\Delta$. (b) The XXZ model with $\Delta=1$ and various $N$. (c) The next-nearest-neighbor model with $N=100$ and various $J_{2}$. (d) The five-state clock model with various $h$ and $N$.}
\label{fig:Figure2}
\end{figure}

For the XXZ model with $\Delta{\neq}0$, the ground state can be computed numerically using the density matrix renormalization group (DMRG)~\cite{White1992}. When the ground state $|\Psi\rangle$ is expressed as a matrix product state (MPS)~\cite{Ostlund1995,Schollwoeck2011}, its crosscap overlap $\langle\mathcal{C}_{\rm latt}|\Psi\rangle$ would be the tensor contraction shown in Fig.~\ref{fig:Figure1}(b). Numerical results at many different $N$'s and $\Delta$'s are displayed in Fig.~\ref{fig:Figure2}. The MPS bond dimension is $2000$ for $N=100$ and $6000$ for $N=400$. The accuracy is very good in a wide range of $\Delta$ for a moderate $N=100$. The relative error of $K$ is of the order $10^{-4}$ or smaller when $\Delta\in[-0.95,0.8]$. As $\Delta$ approaches $1$, clear deviation from exact values is observed [not visible in Fig.~\ref{fig:Figure2}(a) due to its scale] and the relative error reaches $7.20\%$ at $\Delta=1$. This deviation is tentatively attributed to the marginal terms in the low-energy theory at $\Delta=1$~\cite{Affleck1989}, which usually cause strong finite-size effects. It does get smaller as $N$ increases but the convergence is very slow [see Fig.~\ref{fig:Figure2}(b)]. To validate this conjecture, we study the model $H_{\rm next} = \sum^{N}_{j=1} (\mathbf{S}_{j} \cdot \mathbf{S}_{j+1} + J_{2} \mathbf{S}_{j} \cdot \mathbf{S}_{j+2})$ with a next-nearest-neighbor Heisenberg interaction, for which the marginal term disappears when $J_{2}{\approx}0.24$~\cite{Affleck1986a}. The results for this model with $N=100$ are presented in Fig.~\ref{fig:Figure2}(c). As $J_{2}$ increases toward $0.24$, the relative error of $K$ decreases and eventually becomes $0.11\%$. Further analysis based on conformal perturbation theory reveals that finite-size results for the XXZ model with $\Delta=1$ should be fitted as $K=a[\ln(N)]^{-1}+b$~\cite{Append,ZhangYS2024}. This leads to $b=0.510$ using the data points with $N\in[100,400]$ in Fig.~\ref{fig:Figure2}(b).

{\em Spin chains without microscopic U(1) symmetry} --- A fundamental improvement of our method is that microscopic U(1) symmetry is not required. To demonstrate this advantage, we consider the quantum $q$-state clock model~\cite{Elitzur1979,Froehlich1981,Ortiz2012,LiW2015,SunG2019,Berenstein2023}
\begin{align}
H_{q{\rm SC}} = - \sum^{N}_{j=1} ( \sigma^{\dag}_{j} \sigma_{j+1} + \sigma_{j} \sigma^{\dag}_{j+1}) - h \sum^{N}_{j=1} (\tau_{j} + \tau_{j}^{\dag} ) \, ,
\end{align}
where the $\mathbb{Z}_q$ spin operators are defined as $\sigma |\alpha\rangle = |{\rm mod}(\alpha-1+q,q)\rangle$ and $\tau |\alpha\rangle = e^{i2\pi\alpha/q} |\alpha\rangle$ with $\alpha = 1,\ldots,q$. This model is self-dual (SD) at $h=1$. For $q{\geq}5$ and intermediate $h$ (including the SD point), it hosts a Luttinger liquid that is separated by two Kosterlitz–Thouless (KT) transitions at $h_{\rm KT1}$ and $h_{\rm KT2}$ from gapped phases~\cite{Wiegmann1978,Matsuo2006}. The transition points are related to each other by duality as $h_{\rm KT1}h_{\rm KT2}=1$. There is no analytical expression for the Luttinger parameter, but its value at the SD point and the KT transition points was predicted to be $K_{\rm KT1} = q^{2}/8, K_{\rm SD} = q/2, K_{\rm KT2} = 2$~\cite{Ortiz2012,LiZQ2020}. As in previous models, the Ansatz
\begin{align}
|\mathcal{C}_{\mathrm{latt}}\rangle = \prod_{j=1}^{N/2} \sum_{\alpha=1}^{q}|\alpha\rangle_j |\alpha\rangle_{j+N/2} 
\end{align}
is proposed as a crosscap state. Numerical results for the $q=5$ case with $N=50,\cdots,90$ and various $h$ are presented in Fig.~\ref{fig:Figure2}(d). The MPS bond dimension is $1500$ for $N=50$ and $2500$ for $N=90$. At the SD point, finite-size systems are described by CFT with irrelevant perturbations, and the Luttinger parameter has a relative error $0.79\%$ at $N=90$. In the Luttinger liquid regime, the crosscap overlap only exhibits weak dependence on the system size, so the data points for different $N$ almost coincide. For the small and large $h$ phases, the overlap strongly depends on $N$.

{\em Summary and outlook} --- To summarize, we have unveiled that the Luttinger parameter can be extracted from the overlaps between individual eigenstates and the crosscap states. The advantages of our method are corroborated by analytical and numerical results. This method may be applied to check if a trial wave function indeed describes the physics of TLL or other CFTs (see, e.g., Refs.~\cite{Cirac2010,Patil2017}) in the absence of a Hamiltonian. From the theoretical perspective, our results open up an exciting new avenue and many interesting directions are yet to be explored. The crosscap state used here is constructed from a certain type of Bell states. For integrable spin chains, other types of crosscap states have been studied~\cite{Ekman2022}. Do they have clear counterparts in CFTs? An extension of our method to multicomponent Luttinger liquids, such as the fermionic Hubbard model and SU(3) spin chains, is certainly very desirable. Finally, experimental protocols for measuring the crosscap overlap should be pursued. We hope that interesting results on these topics will be reported in the future.

{\em Acknowledgments} --- We are grateful to Meng Cheng for stimulating discussions. B.Y.T. and Y.H.W. are supported by NNSF of China under Grant No.~12174130. Y.S.Z. is supported by the Sino-German (CSC-DAAD) Postdoc Scholarship Program. W.T. is supported by the Research Foundation Flanders (FWO) via grant GOE1520N. L.W. is supported by the NNSF of China under Grants No.~T2225018 and No.~92270107. 

\bibliography{refs}

\clearpage
\onecolumngrid

\setcounter{figure}{0}
\setcounter{table}{0}
\setcounter{equation}{0}
\renewcommand{\thefigure}{A\arabic{figure}}
\renewcommand{\thetable}{A\arabic{table}}
\renewcommand{\theequation}{A\arabic{equation}}

\appendix

\section{Appendix A: Tomonaga-Luttinger liquid and compactified boson conformal field theory}

The bosonized version of the (single-component) Tomonaga-Luttinger liquid is formulated as a free massless compactified boson field theory in (1+1)-dimensional spacetime. This section is devoted to a brief review of the operator formalism of the compactified boson field theory, and we mostly follow the notation used in Refs.~\cite{Francesco-Book,ZhangHC2022}. We assume that the spatial direction is compact with length $L$, and the system admits periodic boundary condition along this direction. Therefore, the spacetime manifold (or ``world sheet''), parametrized by coordinates $(x,t)$, is effectively a cylinder of circumference $L$. The Hamiltonian reads
\begin{align}
    H = \frac{v}{8\pi} \int^{L}_{0} \mathrm{d}x \, \left[ (\partial_{x}\hat{\varphi})^{2} + (\partial_{x}\hat{\theta})^{2} \right],
\label{eq:LuttHami-SM}
\end{align}
where $\hat{\varphi}$ and $\hat{\theta}$ are compactified boson fields, and $v$ is the ``speed of light'' in this theory. Apart from $v$, the theory is characterized by the compactification radius $R$ of $\hat{\varphi}$, which is related to the Luttinger parameter via $K=R^{2}/4$. The field $\hat{\theta}$ is dual to $\hat{\varphi}$ in the sense that $\partial_{x} \hat{\theta}(x,t) = v^{-1} \partial_{t} \hat{\varphi}(x,t)$.

On this manifold, the mode expansion of the $\hat{\varphi}$-field is given by
\begin{align}
    \hat{\varphi}(x,t) = \varphi_{0} + \frac{4\pi}{L} \left( \pi_{0}vt + \tilde{\pi}_{0}x \right) + i \sum_{k \neq 0} \frac{1}{k} \left( a_{k} e^{2\pi ik\frac{x-vt}{L}} + \bar{a}_{k} e^{-2\pi ik\frac{x+vt}{L}} \right),
\label{eq:mode-expan-phi-SM}
\end{align}
where the zero-mode operators satisfy canonical commutation relations (we have formally introduced $\tilde{\varphi}_{0}$ as the operator canonically conjugate to $\tilde{\pi}_{0}$),
\begin{align}
    [\varphi_{0},\pi_{0}] = [\tilde{\varphi}_{0},\tilde{\pi}_{0}] = i,
\label{eq:canonical-commutation-SM}
\end{align}
whilst the other modes generate the Heisenberg algebra
\begin{align}
    [a_{k},a_{k^{\prime}}]=[\bar{a}_{k},\bar{a}_{k^{\prime}}]=k\delta_{k+k^{\prime}}, \quad
    [a_{k},\bar{a}_{k^{\prime}}]=0.
\label{eq:heisenberg-algebra-SM}
\end{align}
The fact the $\hat{\varphi}$-field is compactified with radius $R$ means that the field configurations $\hat{\varphi}$ and $\hat{\varphi} + 2\pi R$ are identified with each other. This implies that the field can have non-trivial winding around the cylinder, i.e., $\hat{\varphi}(x+L, t) \equiv \hat{\varphi}(x, t) - 2\pi mR$ with $m \in \mathbb{Z}$; Eq.~\eqref{eq:mode-expan-phi-SM} then shows that the operator $\tilde{\pi}_{0}$ has eigenvalues $-mR/2$.

Before proceeding, let us remark on the physical significance of the zero-mode operators $\varphi_0$ and $\pi_0$. By definition, the ``canonical momentum'' (density) corresponding to the $\hat{\varphi}$-field is
\begin{align}
    \Pi(x,t) \equiv \frac{1}{4\pi v} \partial_{t} \hat{\varphi}(x, t) = \frac{\pi_0}{L} + \frac{1}{2L} \sum_{k \neq 0} \left( a_{k} e^{2\pi ik\frac{x-vt}{L}} + \bar{a}_{k} e^{-2\pi ik\frac{x+vt}{L}} \right).
\end{align}
Indeed, the canonical equal-time commutation relation, $[\hat{\varphi}(x, 0), \Pi(x^{\prime},0)] = i \delta(x - x^{\prime})$, can readily be verified. The resulting $\mathrm{U}(1)$ conserved charge, $\int_{0}^{L} \mathrm{d}x~\Pi(x,t)$, is precisely $\pi_0$; this indicates that the latter admits an interpretation as the ``centre-of-mass momentum'' of the field. In particular, the single-valuedness of the operator $e^{2\pi i R \pi_0}$ translating around the compactified dimension (in the space of field configurations; not to be confused with the worldsheet) enforces that the eigenvalues of $\pi_0$ are quantized to be $n/R$ with $n \in \mathbb{Z}$. As $\varphi_0$ is conjugate to $\pi_0$, the former is naturally interpreted as the ``centre-of-mass coordinate'' operator of the field.

Next, we recombine these zero-mode operators as follows:
\begin{align}
    Q&=\frac{1}{2}(\varphi_{0}-\tilde{\varphi}_{0}), \quad
    P=\pi_{0}-\tilde{\pi}_{0}; \nonumber\\
    \bar{Q}&=\frac{1}{2}(\varphi_{0}+\tilde{\varphi}_{0}), \quad \bar{P}=\pi_{0}+\tilde{\pi}_{0}.
\end{align}
It is easily verified that these new operators still obey the canonical commutation relations,
\begin{align}
    [Q, P] = [\bar{Q}, \bar{P}] = i.
\end{align}
In terms of this new set of zero-mode operators as well as the ``light-cone'' coordinates $z \equiv -(x-vt)/L$ and $\bar{z} \equiv -(x+vt)/L$, one arrives at the chiral decomposition of the $\hat{\varphi}$-field, $\hat{\varphi}(x,t) = \phi(z) + \bar{\phi}(\bar{z})$, where
\begin{align}
    \phi(z) &= Q + 2\pi Pz + i\sum_{k \neq 0} \frac{1}{k}a_{k}e^{-2\pi ikz}, \nonumber\\
    \bar{\phi}(\bar{z}) &= \bar{Q} - 2\pi\bar{P}\bar{z} + i\sum_{k \neq 0} \frac{1}{k}\bar{a}_{k}e^{2\pi ik\bar{z}}.
\label{eq:mode-expan-SM}
\end{align}
Using the relations $\partial_{x} = -L^{-1} (\partial + \bar{\partial})$ and $\partial_{t} = vL^{-1} (\partial - \bar{\partial})$ with $\partial \equiv \partial_{z}$ and $\bar{\partial} \equiv \partial_{\bar{z}}$, one finds that $\hat{\theta}(x,t) = \bar{\phi}(\bar{z}) - \phi(z)$, we obtain
\begin{align}
    \hat{\theta}(x,t) = \tilde{\varphi}_{0} + \frac{4\pi}{L} \left( \pi_{0}x +\tilde{\pi}_{0}vt \right) - i\sum_{k \neq 0} \frac{1}{k} \left( a_{k} e^{2\pi ik \frac{x-vt}{L}} - \bar{a}_{k}e^{-2\pi ik \frac{x+vt}{L}} \right).
\label{eq:mode-expan-theta-SM}
\end{align}
As $\hat{\theta}(x+L,t) = \hat{\theta}(x,t) + 4\pi\pi_{0}$, the quantization of the eigenvalues of $\pi_{0}$ implies that the $\hat{\theta}$-field is also compactified, for which the radius is $2/R$.

Finally, substituting the mode expansions~\eqref{eq:mode-expan-phi-SM} and~\eqref{eq:mode-expan-theta-SM} into the expression~\eqref{eq:LuttHami-SM} of the Hamiltonian, and making use of~\eqref{eq:heisenberg-algebra-SM} as well as the zeta-function regularization (``$1 + 2 + 3 + \cdots = -1/12$''), one finds after a straightforward computation that
\begin{align}
    H = \frac{2\pi v}{L} \left( L_{0} + \bar{L}_{0} - \frac{1}{12} \right),
\label{eq:CFT-Hamiltonian-SM}
\end{align}
where
\begin{align}
    L_0 &= \frac{1}{2}a_0^2+\sum_{k=1}^{\infty} a_{-k}a_k, \nonumber\\
    \bar{L}_0 &= \frac{1}{2}\bar{a}_0^2+\sum_{k=1}^{\infty} \bar{a}_{-k}\bar{a}_k,
\end{align}
with $a_0 \equiv P,~\bar{a}_0 \equiv \bar{P}$, are the zeroth Virasoro generators. The Virasoro primary states (which are normalized by assumption), $|n,m\rangle$, are labelled by the eigenvalues of $a_0$ and $\bar{a}_0$ as
\begin{align}
    a_{0} |n,m\rangle &= \left( \frac{n}{R} + \frac{1}{2} mR \right) |n,m\rangle, \nonumber\\
    \bar{a}_{0} |n,m\rangle &= \left( \frac{n}{R} - \frac{1}{2} mR \right) |n,m\rangle.
\end{align}
By definition, they are annihilated by the modes $a_{k}$ and $\bar{a}_{k}$ with $k\in\mathbb{Z}^+$; other states in the corresponding conformal tower, which are orthogonal to each other and properly normalized, are constructed by acting modes with negative indices on the primary states:
\begin{align}
    |n,m; \{n_k\},\{\bar{n}_k\}\rangle = \prod_{k=1}^{\infty} \left[\frac{1}{\sqrt{n_k!\bar{n}_k!}}\left(\frac{a_{-k}}{\sqrt{k}}\right)^{n_k}\left(\frac{\bar{a}_{-k}}{\sqrt{k}}\right)^{\bar{n}_k}\right] |n,m\rangle,\quad n_k,\bar{n}_k\in\mathbb{Z}^{+}_{0} \, .
\label{eq:basis-SM}
\end{align}
One obviously has $|n,m;\{0\},\{0\}\rangle \equiv |n,m \rangle$. This basis trivializes the computation of partition functions, as $L_0$ and $\bar{L}_0$ act diagonally:
\begin{align}
    L_0 |n,m;\{n_k\},\{\bar{n}_k\}\rangle &= \left[\frac{1}{2}\left(\frac{n}{R}+\frac{1}{2}mR\right)^2 +\sum_{k=1}^{\infty} kn_k\right] |n,m;\{n_k\},\{\bar{n}_k\}\rangle \, , \nonumber \\
    \bar{L}_0 |n,m;\{n_k\},\{\bar{n}_k\}\rangle &= \left[\frac{1}{2}\left(\frac{n}{R}-\frac{1}{2}mR\right)^2 +\sum_{k=1}^{\infty} k\bar{n}_k\right] |n,m;\{n_k\},\{\bar{n}_k\}\rangle \, .
\label{eq:basis-2-SM}
\end{align}

\section{Appendix B: Determining crosscap states from constraints}

The mode expansion form of the crosscap states $|\mathcal{C}_\pm\rangle$ is given in Eq.~(9) of the main text. Below we derive it in two steps:

i) Substitute the mode expansions of the compactified boson fields~\eqref{eq:mode-expan-SM} into the constraints 
\begin{align}
    e^{i[\hat{\theta}(x) \mp \hat{\theta}(x+L/2)]/R'}|\mathcal{C}_{\pm}\rangle &= |\mathcal{C}_\pm\rangle \, ,   \nonumber \\
    e^{i[\hat{\varphi}(x) \pm \hat{\varphi}(x+L/2)]/R}|\mathcal{C}_{\pm}\rangle &= |\mathcal{C}_\pm\rangle
\label{eq:gluing-SM}
\end{align}
to determine the crosscap states $|\mathcal{C}_\pm\rangle$ up to overall factors $\kappa_{\pm}$.

ii) Compare the cylinder partition function with two crosscap boundaries (``tree-channel''), $Z^{\mathcal{C}_\pm}(L,\beta)=\langle \mathcal{C}_\pm|e^{-\beta H}|\mathcal{C}_\pm\rangle$, with the Klein bottle partition function (``loop channel''), $Z^{\mathcal{K}_\pm} (2\beta, L/2)=\mathrm{Tr}(\Omega_\pm e^{-\beta H})$ to fix the factors $\kappa_{\pm}$.

\textbf{Step i):}

Consider the mode expansions of the compactified boson fields in Eqs.~\eqref{eq:mode-expan-phi-SM} and \eqref{eq:mode-expan-theta-SM} at $t=0$:
\begin{align}
    \hat{\theta}(x) = \tilde{\varphi}_{0} + \frac{4\pi x}{L} \pi_{0} - i\sum_{k \neq 0} \frac{1}{k} \left( a_{k} e^{2\pi ik \frac{x}{L}} - \bar{a}_{k}e^{-2\pi ik \frac{x}{L}} \right), \nonumber \\
    \hat{\varphi}(x) = \varphi_{0} + \frac{4\pi x}{L} \tilde{\pi}_{0} + i \sum_{k \neq 0} \frac{1}{k} \left( a_{k} e^{2\pi ik\frac{x}{L}} + \bar{a}_{k} e^{-2\pi ik\frac{x}{L}} \right).
\end{align}
To analyze the constraints in Eq.~\eqref{eq:gluing-SM}, we consider
\begin{align}
    \hat{\theta}(x) - \hat{\theta}(x+L/2) &= -2\pi \pi_{0} - 2i\sum_{k \neq 0, \, k \; \mathrm{odd}} \frac{1}{k} \left( a_{k} e^{2\pi ik \frac{x}{L}} - \bar{a}_{k}e^{-2\pi ik \frac{x}{L}} \right), \nonumber \\
    \hat{\varphi}(x) +  \hat{\varphi}(x+L/2) &= 2\varphi_{0} + 4\pi \tilde{\pi}_{0} \left(\frac{2x}{L}+\frac{1}{2}\right) + 2i \sum_{k \neq 0, \, k \; \mathrm{even}} \frac{1}{k} \left( a_{k} e^{2\pi ik\frac{x}{L}} + \bar{a}_{k} e^{-2\pi ik\frac{x}{L}} \right),
\end{align}
and
\begin{align}
    \hat{\theta}(x) + \hat{\theta}(x+L/2) &= 2\tilde{\varphi}_{0} + 4\pi \pi_{0} \left(\frac{2x}{L}+\frac{1}{2}\right) - 2i\sum_{k \neq 0, \, k \; \mathrm{even}} \frac{1}{k} \left( a_{k} e^{2\pi ik \frac{x}{L}} - \bar{a}_{k}e^{-2\pi ik \frac{x}{L}} \right), \nonumber \\
    \hat{\varphi}(x) -  \hat{\varphi}(x+L/2) &= -2\pi \tilde{\pi}_{0} + 2i \sum_{k \neq 0, \, k \; \mathrm{odd}} \frac{1}{k} \left( a_{k} e^{2\pi ik\frac{x}{L}} + \bar{a}_{k} e^{-2\pi ik\frac{x}{L}} \right).
\end{align}
By inserting these into Eq.~\eqref{eq:gluing-SM} and noting that the constraints~\eqref{eq:gluing-SM} are valid for all $x \in [0,L]$, it is then obvious that the oscillatory part should satisfy
\begin{align}
    \left[a_k\mp (-1)^k\bar{a}_{-k}\right]|\mathcal{C}_{\pm}\rangle &=0,\quad k\neq 0 \, ,
\end{align}
which fixes the form of $|\mathcal{C}_\pm\rangle$ to be
\begin{align}
    |\mathcal{C}_{\pm}\rangle & \propto \exp\left[\pm\sum_{k=1}^\infty\frac{(-1)^k}{k}a_{-k}\bar{a}_{-k}\right]|\mathcal{C}^0_\pm\rangle \, .
\label{eq:cross-SM}
\end{align}
Here $|\mathcal{C}^0_\pm\rangle = \sum_{n,m\in\mathbb{Z}} c^{\pm}_{n,m}|n,m\rangle$ encode the zero mode information of the crosscap states, where $c^{\pm}_{n,m}$ are the superposition coefficients that should be determined by using the zero mode constraints from Eq.~\eqref{eq:gluing-SM}:
\begin{align}
    e^{- 2\pi i \pi_{0}/R^{\prime}}|\mathcal{C}^0_+\rangle =|\mathcal{C}^0_+\rangle \, ,\nonumber \\
    e^{2i\varphi_0/R} e^{4\pi i\left(\frac{2x}{L}+\frac{1}{2}\right)\tilde{\pi}_0/R}|\mathcal{C}^0_+\rangle =|\mathcal{C}^0_+\rangle \, ,
\label{eq:gluing-1-SM}
\end{align}
and
\begin{align}
    e^{2i\tilde{\varphi}_{0}/R'} e^{4\pi i  \left(\frac{2x}{L}+\frac{1}{2}\right)\pi_{0}/R'}|\mathcal{C}^0_-\rangle =|\mathcal{C}^0_-\rangle \, , \nonumber \\
    e^{-2\pi i \tilde{\pi}_{0}/R}|\mathcal{C}^0_-\rangle =|\mathcal{C}^0_-\rangle \, .
\label{eq:gluing-2-SM}
\end{align}
For $|\mathcal{C}^0_+\rangle = \sum_{n,m\in\mathbb{Z}} c^{+}_{n,m}|n,m\rangle$, the first equation in Eq.~\eqref{eq:gluing-1-SM} requires $c^+_{2n+1,m}=0$, since $e^{-2\pi i\pi_0/R'}|n,m\rangle =(-1)^n |n,m\rangle$ (note that $\pi_0|n,m\rangle=\tfrac{n}{R}|n,m\rangle$ and $R^{\prime} = 2/R$); the second equation in Eq.~\eqref{eq:gluing-1-SM} requires  $c^+_{n,m\neq 0}=0$ and $c^+_{2n,0}=c^+_{0,0} \; \forall n\in\mathbb{Z}$, since $e^{4\pi i(\frac{2x}{L}+\frac{1}{2})\tilde{\pi}_0/R}|n,m\rangle =e^{2\pi i(\frac{2x}{L}+\frac{1}{2})m}|n,m\rangle$ and $ e^{2i\varphi_0/R}|n,m\rangle = |n+2,m\rangle$ (note that $\tilde{\pi}_0|n,m\rangle=-\tfrac{1}{2}mR|n,m\rangle$ and $[\varphi_0,\pi_0]=i$). Thus, we obtain 
\begin{align}
    |\mathcal{C}^0_+\rangle \propto \sum_{n\in\mathbb{Z}}|2n,0\rangle \, .
\end{align}
The determination of $|\mathcal{C}^0_-\rangle$ is similar and will not be repeated. It is given by
\begin{align}
    |\mathcal{C}^0_-\rangle \propto \sum_{m\in\mathbb{Z}}|0,2m\rangle\,.
\end{align}
Altogether, the constraints in Eq.~\eqref{eq:gluing-SM} determine the mode expansion form of the crosscap states
\begin{align}
     |\mathcal{C}_{+}\rangle &=\kappa_+ \exp\left[\sum_{k=1}^\infty \frac{(-1)^k}{k}a_{-k}\bar{a}_{-k}\right] \sum_{n\in\mathbb{Z}} |2n,0\rangle \, ,\nonumber \\
    |\mathcal{C}_{-}\rangle &=\kappa_- \exp\left[-\sum_{k=1}^\infty \frac{(-1)^k}{k}a_{-k}\bar{a}_{-k}\right] \sum_{m\in\mathbb{Z}} |0,2m\rangle
\label{eq:cross-2-SM}
\end{align}
up to overall factors $\kappa_{\pm}$. Without loss of generality, we choose \textbf{}$\kappa_{\pm}>0$ below.

\textbf{Step ii):}

To fix the factors $\kappa_\pm$ in Eq.~\eqref{eq:cross-2-SM}, we first calculate the (``tree channel'') cylinder partition function with crosscap boundary states 
\begin{align}
    Z^{\mathcal{C}_\pm}(L,\beta ) =\langle \mathcal{C}_\pm |e^{-\beta H}|\mathcal{C}_\pm\rangle \, ,
\end{align}
where the Hamiltonian $H$ is given in Eq.~\eqref{eq:CFT-Hamiltonian-SM}. The crosscap states in Eq.~\eqref{eq:cross-2-SM} can be expanded in the orthonormal basis states in Eq.~\eqref{eq:basis-SM} as
\begin{align}
    |\mathcal{C}_+\rangle &=\kappa_+ \prod_{k=1}^\infty\sum_{n_k=0}^\infty \frac{1}{n_k!}\left[\frac{(-1)^k a_{-k}\bar{a}_{-k}}{k}\right]^{n_k} \sum_{n\in\mathbb{Z}} |2n,0\rangle \nonumber \\
    &=\kappa_+ \sum_{n\in\mathbb{Z}} \sum_{n_1,n_2,\ldots =0}^{\infty} (-1)^{\sum\limits_{k=1}^\infty k n_k} |2n,0; \{n_k\},\{n_k\}\rangle \, ,
\end{align}
and 
\begin{align}
    |\mathcal{C}_-\rangle &=\kappa_- \prod_{k=1}^\infty\sum_{n_k=0}^\infty\frac{1}{n_k!}\left[\frac{(-1)^{k+1} a_{-k}\bar{a}_{-k}}{k}\right]^{n_k} \sum_{m\in\mathbb{Z}} |0,2m\rangle \nonumber \\
    &=\kappa_- \sum_{m\in\mathbb{Z}} \sum_{n_1,n_2,\ldots =0}^{\infty} (-1)^{\sum\limits_{k=1}^\infty (k+1) n_k} |0,2m;\{n_k\};\{n_k\}\rangle \, .
\end{align}
As $|n,m;\{n_k\},\{\bar{n}_k\}\rangle$ is the Hamiltonian eigenbasis [Eq.~\eqref{eq:basis-2-SM}], $Z^{\mathcal{C}_{\pm}}(L,\beta)$ are easy to calculate
\begin{align}
    Z^{\mathcal{C}_+}(L,\beta) &= \kappa_+^2 \sum_{n\in\mathbb{Z}} \sum_{n_1,n_2,\ldots =0}^{\infty} \langle 2n,0; \{n_k\},\{n_k\}|q^{L_0+\bar{L}_0-1/12} |2n,0;\{n_k\},\{n_k\}\rangle \nonumber \\
    &=\kappa_+^2\; q^{-1/12} \sum_{n\in\mathbb{Z}} q^{{R'}^2n^2} \sum_{n_1,n_2,\ldots =0}^{\infty} q^{2\sum_{k=1}^\infty k n_k} \nonumber \\
    &= \kappa_+^2 \sum_{n\in\mathbb{Z}} q^{{R'}^2n^2} \cdot q^{-1/12}\prod_{k=1}^\infty\frac{1}{1-q^{2k}}\nonumber\\
    &=\kappa_+^2 \, \frac{\vartheta_3(2{R'}^2\tau)}{\eta (2\tau)} \, ,
\end{align}
and similarly
\begin{align}
    Z^{\mathcal{C}_-}(L,\beta) = \kappa_-^2 \, \frac{\vartheta_3(2{R}^2\tau)}{\eta (2\tau)}\,,
\end{align}
where $\tau = iv\beta/L$ is the modular parameter and $q = e^{2\pi i\tau}$. The results are expressed in terms of the Dedekind eta function $\eta(\tau) = q^{1/24}\prod_{k=1}^{\infty} (1-q^k)$ and one of the standard Jacobi's theta functions $\vartheta_{3}(\tau) = \sum_{k \in\mathbb{Z}} q^{k^{2}/2}$. Under the modular $S$ transformation ($\tau \rightarrow -1/\tau$), Dedekind eta function and Jacobi's theta function transform as $\sqrt{-i\tau}\vartheta_3(\tau)=\vartheta_3(-1/\tau)$ and $\sqrt{-i\tau}\eta(\tau)=\eta(-1/\tau)$, respectively. This allows us to rewrite $Z^{\mathcal{C}_{\pm}}$ as
\begin{align}
    Z^{\mathcal{C}_+}(L,\beta) &=\kappa_+^2 \, \frac{1}{R'} \, \frac{\vartheta_3(-1/2{R'}^2\tau)}{\eta (-1/2\tau)} \, , \nonumber \\
    Z^{\mathcal{C}_-}(L,\beta) &=\kappa_-^2 \, \frac{1}{R} \, \frac{\vartheta_3(-1/2{R'}^2\tau)}{\eta (-1/2\tau)} \, .
\label{eq:Z-crosscap-SM}
\end{align}

Now we turn to the (``loop channel'') Klein bottle partition function~\cite{TangW2019}
\begin{align}
    Z^{\mathcal{K}_\pm}(2\beta,L/2) = \mathrm{Tr}\left(\Omega_\pm e^{-\tilde{H}L/2}\right) ,
\end{align}
where $\tilde{H} = \frac{2\pi}{\tilde{L}}(L_0+\bar{L}_0-\frac{1}{12})$ is the compactified boson Hamiltonian [c.f.~Eqs.~\eqref{eq:LuttHami-SM} and \eqref{eq:CFT-Hamiltonian-SM}] defined on a circle (in the imaginary time direction) with circumference $\tilde{L}=2v\beta$, which generates evolution along the spatial direction (it can be viewed as the generator of the quantum transfer matrix). The action of the ``reflection operator'' $\Omega_\pm$ on oscillatory modes is given by $\Omega^{-1}_{\pm}a_k\Omega_{\pm} = \mp \bar{a}_k$ and $\Omega^{-1}_{\pm}\bar{a}_k\Omega_{\pm} = \mp a_k$ for $k\neq 0$. Applying them to Virasoro primary states gives
\begin{align}
    \Omega_+ |n,m\rangle &= |-n,m\rangle \, , \nonumber \\
    \Omega_- |n,m\rangle &= |n,-m\rangle \, .
\end{align}
Evaluating the Klein bottle partition function $Z^{\mathcal{K}_\pm}$ in the Hamiltonian eigenbasis [Eq.~\eqref{eq:basis-2-SM}] and noticing that only ``left-right-symmetric'' states $|0,m;\{n_k\},\{n_k\}\rangle$ ($|n,0;\{n_k\},\{n_k\}\rangle$) contribute to $Z^{\mathcal{K}_+}$ ($Z^{\mathcal{K}_-}$), we obtain
\begin{align}
    Z^{\mathcal{K}_+}(2\beta,L/2) &= \sum_{m\in\mathbb{Z}} \sum_{n_1,n_2,\ldots =0}^{\infty} \langle 0,m; \{n_k\},\{n_k\}|\tilde{q}^{2L_0-1/12}|0,m;\{n_k\},\{n_k\} \rangle \nonumber \\
    &=\tilde{q}^{-1/12}\sum_{m\in\mathbb{Z}}\tilde{q}^{m^2/{R'}^2} \sum_{n_1,n_2,\ldots =0}^{\infty}  \tilde{q}^{2\sum_{k=1}^\infty k n_k} \nonumber \\
    &=\frac{\vartheta_3(-1/2{R'}^2\tau)}{\eta (-1/2\tau)} \, ,
\label{eq:Z-KB1-SM}
\end{align}
and similarly
\begin{align}
    Z^{\mathcal{K}_-}(2\beta,L/2) &= \sum_{n\in\mathbb{Z}} \sum_{n_1,n_2,\ldots =0}^{\infty}  \langle n,0;\{n_k\},\{n_k\}|\tilde{q}^{2L_0-1/12}|n,0;\{n_k\},\{n_k\}\rangle \nonumber \\
    &=\frac{\vartheta_3(-1/2{R}^2\tau)}{\eta (-1/2\tau)}
\label{eq:Z-KB2-SM}
\end{align}
with $\tilde{q}=e^{-\pi L/2 v\beta}= e^{-\pi i/2\tau}$. 

Considering the spacetime symmetry, the CFT partition function evaluated in tree and loop channels should be equal (``loop channel-tree channel equivalence''), so we expect
\begin{align}
    Z^{\mathcal{C}_\pm}(L,\beta) = Z^{\mathcal{K}_\pm}(2\beta,L/2) \, .
\end{align}
Comparing Eqs.~\eqref{eq:Z-crosscap-SM}, \eqref{eq:Z-KB1-SM} and \eqref{eq:Z-KB2-SM}, the factors in the crosscap states~\eqref{eq:cross-2-SM} are fixed to be
\begin{align}
    \kappa_+ &= \sqrt{R^{\prime}} \, , \nonumber\\
    \kappa_- &= \sqrt{R} \, .
\end{align}

\section{Appendix C: Crosscap state overlap of Jastrow wave functions}

In this section, we calculate the overlap between the lattice crosscap state
\begin{align}
    |\mathcal{C}_{\mathrm{latt}}\rangle =\prod_{j=1}^{N/2}\left(|\uparrow\rangle_j|\uparrow\rangle_{j+N/2}+|\downarrow\rangle_j|\downarrow\rangle_{j+N/2}\right)
\label{eq:cross-latt-SM}
\end{align}
and the Jastrow wave function
\begin{align}
    |\Psi^{\lambda}(M)\rangle = \sum_{1\leq x_1<\cdots < x_M\leq N} \Psi^\lambda (x_1,\ldots,x_M) \sigma_{x_1}^+\cdots \sigma_{x_M}^+|\downarrow\cdots\downarrow\rangle
\label{eq:Jastrow-1-SM}
\end{align}
with
\begin{align}
  \Psi^{\lambda} (x_1,\ldots,x_M) = \mathcal{N}_{N,M,\lambda} \, (-1)^{\sum_{l=1}^{M} x_l} \prod_{1 \leq j<k \leq M} 
  \left\vert\sin\frac{\pi(x_j-x_k)}{N}\right\vert^\lambda
\label{eq:Jastrow-2-SM}
\end{align}
for $\mathrm{mod}(N,4)=0$. The number of hardcore bosons, $M$, is taken to be even. For $\lambda = 1$ ($\lambda = 2$), $|\Psi^{\lambda}(M)\rangle$ is an exact eigenstate of the spin-1/2 XY (Haldane-Shastry) chain. $\mathcal{N}_{N,M,\lambda}$ ensures normalization of $|\Psi^{\lambda}(M)\rangle$ and is chosen to be positive.

To calculate the overlap $\langle\mathcal{C}_{\mathrm{latt}}|\Psi^\lambda (M)\rangle$, we first introduce a unitary $U = \exp\left[ i\tfrac{\pi}{2}\sum_{j:\mathrm{odd}}(\sigma^z_j + 1)\right]$ to remove the Marshall sign $(-1)^{\sum_{l=1}^{M} x_l}$ in $|\Psi^\lambda (M)\rangle$:
\begin{align}
    |\tilde{\Psi}^\lambda (M)\rangle = U |\Psi^{\lambda} (M)\rangle
\label{eq:XY-HS-Jastrow-SM}
\end{align}
with
\begin{align}
  \tilde{\Psi}^{\lambda} (x_1,\ldots,x_M) &= \mathcal{N}_{N,M,\lambda} \, \prod_{1 \leq j<k \leq M} 
  \left\vert\sin\frac{\pi(x_j-x_k)}{N}\right\vert^{\lambda} \, \nonumber \\
  &= \mathcal{N}_{N,M,\lambda} \cdot 2^{-\lambda M(M-1)/2}  \prod_{1 \leq j<k \leq M} 
  \left\vert z_{x_j} - z_{x_k} \right\vert^{\lambda}  \, \nonumber \\
  &= \tilde{\mathcal{N}}_{N,M,\lambda} \, \prod_{1 \leq j<k \leq M} 
  \left\vert z_{x_j} - z_{x_k} \right\vert^{\lambda}  \, ,
\end{align}
where $z_{x_j}=\exp (i\frac{2\pi}{N}x_j)$ and $\tilde{\mathcal{N}}_{N,M,\lambda} = 2^{-\lambda M(M-1)/2}\mathcal{N}_{N,M,\lambda}$.

For $\mathrm{mod}(N,4)=0$, it is obvious that the lattice crosscap state $|\mathcal{C}_{\mathrm{latt}}\rangle$ in Eq.~\eqref{eq:cross-latt-SM} is invariant under the action of $U$, $U |\mathcal{C}_{\mathrm{latt}}\rangle = |\mathcal{C}_{\mathrm{latt}}\rangle$. The crosscap state overlap is then rewritten as
\begin{align}
    \langle\mathcal{C}_{\mathrm{latt}}|\Psi^\lambda (M)\rangle = \langle\mathcal{C}_{\mathrm{latt}}|U^{\dag} U |\Psi^\lambda (M)\rangle = \langle\mathcal{C}_{\mathrm{latt}}|\tilde{\Psi}^\lambda (M)\rangle \, .
\end{align}
For even $M$, the overlap $\langle\mathcal{C}_{\mathrm{latt}}|\tilde{\Psi}^\lambda (M)\rangle$ can be simplified as follows:
\begin{align}
\langle \mathcal{C}_{\mathrm{latt}}|\tilde{\Psi}^{\lambda}(M)\rangle &=\tilde{\mathcal{N}}_{N,M,\lambda}\sum_{1\leq x_1<\cdots<x_{M/2}\leq N/2}\Psi^{\lambda}(x_1,\ldots,x_{M/2},x_1+N/2,\ldots,x_{M/2}+N/2) \nonumber \\
&=\tilde{\mathcal{N}}_{N,M,\lambda} \sum_{1\leq x_1<\cdots<x_{M/2}\leq N/2}\,\prod_{1\leq j < k \leq M/2}\left\vert(z_{x_j}-z_{x_k})(z_{x_{j+N/2}}-z_{x_{k+N/2}})\right\vert^{\lambda} \cdot \prod_{1\leq j,k\leq M/2}\left\vert z_{x_j}-z_{x_{k+N/2}}\right\vert^{\lambda} \nonumber \\
&=\tilde{\mathcal{N}}_{N,M,\lambda} \sum_{1\leq x_1<\cdots<x_{M/2}\leq N/2}\,\prod_{1\leq j < k \leq M/2}\left\vert z_{x_j}-z_{x_k} \right\vert^{2\lambda} \cdot \prod_{1\leq j<k\leq M/2}\left\vert z_{x_j}+z_{x_k}\right\vert^{2\lambda} \cdot 2^{\lambda M/2} \nonumber \\
&=2^{\lambda M/2} \tilde{\mathcal{N}}_{N,M,\lambda} \sum_{1\leq x_1<\cdots<x_{M/2}\leq N/2}\,\prod_{1\leq j < k \leq M/2}\left\vert z^2_{x_j}-z^2_{x_k} \right\vert^{2\lambda} \nonumber \\
&=2^{\lambda M/2} \tilde{\mathcal{N}}_{N,M,\lambda} \cdot \frac{1}{\vert\tilde{\mathcal{N}}_{N/2,M/2,\lambda}\vert^2} \, .
\label{eq:crosscap-overlap-SM}
\end{align}
For the last step, we have used $\sum_{1\leq x_1<\ldots<x_{M/2}\leq N/2}\, \prod_{1\leq j < k \leq M/2}|z_{x_j}^2-z_{x_k}^2|^2 =|\tilde{\mathcal{N}}_{N/2,M/2,\lambda}|^{-2}$, which is the normalization condition of $|\tilde{\Psi}^{\lambda}(M/2)\rangle$ defined on a chain with length $N/2$. Remarkably, the crosscap overlap for the Jastrow wave function $|\Psi^{\lambda}(M)\rangle$ defined in Eqs.~\eqref{eq:Jastrow-1-SM} and \eqref{eq:Jastrow-2-SM} is almost solely determined by the normalization factor. Although this interesting result applies to generic $\lambda$, analytical solutions for the normalization factor $\tilde{\mathcal{N}}_{N,M,\lambda}$ are available only for $\lambda=1$ and $2$ (see, e.g., Ref.~\cite{Kuramoto-Book}). For completeness, we provide a self-contained derivation of $\tilde{\mathcal{N}}_{N,M,\lambda}$ for $\lambda=1$ and $2$ below. 

\subsection{XY chain}
For the XY chain with $\lambda=1$, we use the normalization condition $\langle \tilde{\Psi}^{\lambda=1}(M) | \tilde{\Psi}^{\lambda=1}(M) \rangle = 1$ to obtain
\begin{align}
    |\tilde{\mathcal{N}}_{N,M,\lambda=1}|^{-2}  &=\sum_{1\leq x_1<\cdots < x_M\leq N}|\Psi^{\lambda =1} (x_1,\ldots,x_M)|^2\nonumber\\
    &=\sum_{1\leq x_1<\cdots < x_M\leq N}\prod_{1\leq j<k\leq M}|z_{x_j}-z_{x_k}|^2\nonumber\\
    &=\frac{1}{M!}\sum_{1\leq x_1,\ldots,x_M\leq N} \prod_{1\leq j<k \leq M}|z_{x_j}-z_{x_k}|^2\,.
\end{align}
The Jastrow factor $\prod_{1\leq i< j\leq M}(z_{x_j}-z_{x_i})$ can be written as a Vandermonde determinant 
\begin{align}
     \prod_{1\leq j<k\leq M}(z_{x_k}-z_{x_j}) =\mathrm{det}\begin{pmatrix}1 & z_{x_1} &\cdots& z_{x_1}^{M-1}\\1 & z_{x_2} & \cdots & z_{x_2}^{M-1}\\\vdots & \vdots & \ddots & \vdots \\ 1 & z_{x_M} & \cdots & z_{x_M}^{M-1}\end{pmatrix}\equiv \mathrm{det}(\mathbb{V}) \, ,
\end{align}
so we obtain
\begin{align}
    |\tilde{\mathcal{N}}_{N,M,\lambda=1}|^{-2} &=
    \frac{1}{M!}\sum_{1\leq x_1,\ldots,x_M\leq N} \mathrm{det}(\mathbb{V}) \; \mathrm{det}(\mathbb{V^*}) \nonumber \\
    &=\frac{1}{M!}\sum_{1\leq x_1,\ldots, x_M\leq N} \sum_{\sigma\in S_M} \mathrm{sgn}(\sigma) z_{x_1}^{\sigma (1)-1}\cdots z_{x_M}^{\sigma (M)-1}
    \sum_{\tau\in S_M} \mathrm{sgn}(\tau) {(z_{x_1}^*)}^{\tau (1)-1}\cdots {(z_{x_M}^*)}^{\tau (M)-1}\nonumber\\
    &=\frac{1}{M!}\sum_{\sigma\in S_M}\sum_{\tau\in S_M} \mathrm{sgn}(\sigma) \, \mathrm{sgn}(\tau) \left(\sum_{x_1=1}^N z_{x_1}^{\sigma (1)-\tau (1)}\right)\cdots \left(\sum_{x_M=1}^N z_{x_M}^{\sigma (M)-\tau (M)}\right)\nonumber\\
    &=\frac{1}{M!}\sum_{\sigma\in S_M}\sum_{\tau\in S_M} \mathrm{sgn}(\sigma) \, \mathrm{sgn}(\tau) \, \left( N\delta_{\sigma(1),\tau(1)} \right) \cdots \left(N\delta_{\sigma(M),\tau(M)}\right) \nonumber \\
    &=\frac{1}{M!}\sum_{\sigma\in S_M}N^M \nonumber \\
    &= N^M \, ,
\end{align}
where we have used the definition of the determinant [$S_M$ is the permutation of $\{1,2,\ldots,M\}$ and $\mathrm{sgn}(\sigma)$ denotes the signature of the permutation $\sigma$] and the identity $\sum_{x=1}^N z_x^{k}=N\delta_{k,0}$ for $k\in\mathbb{Z}$. Thus, we obtain the normalization factor $\tilde{\mathcal{N}}_{N,M,\lambda=1} = N^{-M/2}$. By using this result and Eq.~\eqref{eq:crosscap-overlap-SM}, we arrive at
\begin{align}
    \langle \mathcal{C}_{\mathrm{latt}}|\Psi^{\lambda=1}(M)\rangle
    =2^{M/2} \cdot N^{-M/2} \cdot \frac{1}{[(N/2)^{-M/4}]^2} 
    =1 \, .
\end{align}

\subsection{Haldane-Shastry chain}

We now turn to the Haldane-Shastry chain ($\lambda=2$) and determine the normalization factor $\tilde{\mathcal{N}}_{N,M,\lambda=2}$. Using the normalization condition $\langle \tilde{\Psi}^{\lambda=2}(M) | \tilde{\Psi}^{\lambda=2}(M) \rangle = 1$, we obtain
\begin{align}
    |\tilde{\mathcal{N}}_{N,M,\lambda=2}|^{-2}  &=\sum_{1\leq x_1< \cdots <x_M\leq N}|\Psi^{\lambda=2}(x_1,\ldots,x_M)|^2 \nonumber \\
    &=\frac{1}{M!}\sum_{1\leq x_1,\ldots,x_M\leq N}\prod_{1\leq j<k \leq M}|z_{x_j}-z_{x_k}|^4 \nonumber \\
    &=\frac{1}{M!}\sum_{1\leq x_1,\ldots,x_M\leq N}\prod_{1\leq j<k \leq M}(z_{x_j}-z_{x_k})^4\frac{1}{z_{x_j}^2 z_{x_k}^2} \nonumber \\
    &=\frac{1}{M!}\sum_{1\leq x_1,\ldots,x_M\leq N}\prod_{i=1}^M z_{x_i}^{-2(M-1)} \prod_{1\leq j<k \leq M}(z_{x_j}-z_{x_k})^4 \, .
\label{eq:norm-2-SM}
\end{align}
The product $\prod_{1\leq j<k\leq M}(z_{x_j}-z_{x_k})^4$ also has a determinant representation called ``confluent alternant''~\cite{Kuramoto-Book}:
\begin{align}
    \prod_{1\leq j<k\leq M}(z_{x_j}-z_{x_k})^4 =\mathrm{det}\begin{pmatrix}
        1 & z_{x_1} & z_{x_1}^2 & z_{x_1}^3 &\cdots & z_{x_1}^{2M-1}\\ 0 & 1 & 2z_{x_1} & 3z_{x_1}^2 & \cdots & (2M-1)z_{x_1}^{2M-2}\\ 1 & z_{x_2} & z_{x_1}^2 & z_{x_1}^3  &\cdots & z_{x_2}^{2M-1}\\
        \vdots & \vdots & \vdots & \vdots & \ddots & \vdots\\
        1 & z_{x_M} & z_{x_M}^2 & z_{x_M}^3 & \cdots & z_{x_M}^{2M-1}\\
        0 & 1 & 2z_{x_M} & 3z_{x_M}^2 & \cdots & (2M-1)z_{x_M}^{2M-2}
    \end{pmatrix}\equiv \mathrm{det}(\mathbb{M}) \, .
\end{align}
After expanding the ``confluent alternant'' $\mathrm{det}(\mathbb{M})$ as a summation of permutation terms (definition of the determinant), the last term in Eq.~\eqref{eq:norm-2-SM} can be calculated as follows:
\begin{align}
    &\phantom{=} \; \sum_{1\leq x_1,\ldots,x_M\leq N}\prod_{i=1}^M z_{x_i}^{-2(M-1)}\prod_{1\leq j<k \leq M}(z_{x_j}-z_{x_k})^4 \nonumber \\
    &=\sum_{1\leq x_1,\ldots,x_M\leq N}\prod_{i=1}^M z_{x_i}^{-2(M-1)} \; \mathrm{det}(\mathbb{M}) \nonumber\\
    &=\sum_{1\leq x_1,\ldots,x_M\leq N}\prod_{i=1}^M z_{x_i}^{-2(M-1)}\sum_{\sigma\in S_{2M}} \mathrm{sgn}(\sigma) \, z_{x_1}^{\sigma (1)-1}(\sigma (2)-1) z_{x_1}^{\sigma (2)-2}\cdots z_{x_M}^{\sigma (2M-1)-1}(\sigma (2M)-1) z_{x_M}^{\sigma (2M)-2}\nonumber\\
    &=\sum_{\sigma\in S_{2M}} \mathrm{sgn}(\sigma) \prod_{j=1}^M (\sigma (2j)-1) \prod_{j=1}^M \left(\sum_{x_j=1}^N z_{x_j}^{\sigma (2j-1)+\sigma (2j)-(2M+1)}\right) \nonumber\\
    &=N^M \sum_{\sigma\in S_{2M}} \mathrm{sgn}(\sigma) \prod_{j=1}^M (\sigma (2j)-1) \prod_{j=1}^M\delta_{\sigma (2j-1)+\sigma (2j),2M+1} \, .
\end{align}
To evaluate the sum, one observes that nonvanishing contributions come from the permutation elements satisfying $\sigma(2j-1) + \sigma(2j) = 2M+1$. All such permutation elements are generated from $\sigma = \{1,2M,2,2M-1,\ldots, M,M+1\}$ [which has $\mathrm{sgn}(\sigma)=+1$] by using two types of permutations: i) $\sigma(2j-1) \leftrightarrow \sigma(2j)$ [which has $\mathrm{sgn}(\sigma)=-1$] and ii) $\{\sigma(2j-1),\sigma(2j)\} \leftrightarrow \{\sigma(2l-1),\sigma(2l)\}$ [which has $\mathrm{sgn}(\sigma)=+1$]. Making use of the first type of permutation, the sum can be restricted to $\sigma(2j-1)<\sigma(2j) \; \forall j=1,\ldots,M$ and is then reduced to a simple sum over $\{\sigma(1),\sigma(3),\ldots,\sigma(2M-1) \} \in S_M$:
\begin{align}
&\phantom{=} \;  \sum_{\sigma\in S_{2M}} \mathrm{sgn}(\sigma) \prod_{j=1}^M (\sigma (2j)-1) \prod_{j=1}^M\delta_{\sigma (2j-1)+\sigma (2j),2M+1}  \nonumber \\
&=  \sum_{\sigma\in S_{2M}, \, \sigma(2j-1)<\sigma(2j)} \mathrm{sgn}(\sigma) \prod_{j=1}^M [(\sigma (2j)-1) - (\sigma (2j-1)-1)]  \prod_{j=1}^M\delta_{\sigma (2j-1)+\sigma (2j),2M+1}   \nonumber \\
&=  \sum_{\sigma\in S_{2M}, \, \sigma(2j-1)<\sigma(2j)} \mathrm{sgn}(\sigma) \prod_{j=1}^M [\sigma (2j) - \sigma (2j-1)]  \prod_{j=1}^M\delta_{\sigma (2j-1)+\sigma (2j),2M+1}   \nonumber \\
&=  \sum_{\{\sigma(1),\sigma(3),\ldots,\sigma(2M-1) \} \in S_M} (+1) \prod_{j=1}^M [2M+1 - 2 \, \sigma (2j-1)]   \nonumber \\
&=  M! \; (2M-1)(2M-3)\cdots 3 \cdot 1   \nonumber \\
&=  M! \; \frac{(2M)!}{(2M)(2M-2)\cdots 2}   \nonumber \\
&= \frac{(2M)!}{2^M} \, .
\end{align}

Combining the above results, we obtain
\begin{align}
    |\tilde{\mathcal{N}}_{N,M,\lambda=2}|^{-2} = \frac{1}{M!} \cdot N^M \cdot \frac{(2M)!}{2^M} = \left(\frac{N}{2}\right)^{M}\frac{(2M)!}{M!} \, .
\end{align}
and $\tilde{\mathcal{N}}_{N,M,\lambda=2} = \left(\tfrac{2}{N}\right)^{M/2}\sqrt{\tfrac{M!}{(2M)!}}$. By using this result and Eq.~\eqref{eq:crosscap-overlap-SM}, the crosscap state overlap for the Haldane-Shastry chain is written as
\begin{align}
    \langle \mathcal{C}_{\mathrm{latt}}|\Psi^{\lambda=2}(M)\rangle &= 2^{M} \cdot \left(\frac{2}{N}\right)^{M/2}\sqrt{\frac{M!}{(2M)!}} \cdot \left(\frac{N}{4}\right)^{M/2}\frac{M!}{(M/2)!} \nonumber \\
    &= \frac{2^{M/2}(M!)^{3/2}}{(M/2)!\sqrt{(2M)!}} \, .
\end{align}
Using the Stirling approximation $M!= \sqrt{2\pi M}\left(\tfrac{M}{e}\right)^M \left(1+\tfrac{1}{12 M} +\mathcal{O}(M^{-2})\right)$ (valid for $M$ large), we obtain
\begin{align}
\langle \mathcal{C}_{\mathrm{latt}}|\Psi^{\lambda=2}(M)\rangle = 2^{\frac{1}{4}}\left(1-\frac{1}{16M}+\mathcal{O}\left(\frac{1}{M^2}\right)\right)=2^{\frac{1}{4}}+\mathcal{O}(M^{-1}) .
\end{align}

\section{Appendix D: Logarithmic correction to the crosscap overlap at the Heisenberg point}

In this section, we provide a field theoretical analysis for understanding the finite-size scaling behavior of the overlap between the (lattice) crosscap state $|\mathcal{C}_{\mathrm{latt}}\rangle$ and the ground state of the spin-1/2 Heisenberg chain. In particular, we focus on the influence of the marginally irrelevant term.

The effective theory of the spin-1/2 Heisenberg chain is the $\mathrm{SU}(2)_1$ Wess-Zumino-Witten (WZW) CFT, with an SU(2) symmetric \emph{marginal} perturbation:
\begin{align}
    H = H_0 + H_1
    \label{eq:eff-ham-SM}
\end{align}
with
\begin{align}
    H_0 &= \frac{v}{2  \pi}\int_0^L \mathrm{d}x \left[\frac{1}{3}:\mathbf{J}\cdot \mathbf{J}:(x) + \frac{1}{3}:\bar{\mathbf{J}}\cdot \bar{\mathbf{J}}: (x)\right] - \frac{2\pi v}{L}\frac{c}{12} \, ,\nonumber \\
    H_1 &= - vg \int_0^L \mathrm{d}x \;\frac{2 }{\sqrt{3}}:\mathbf{J}\cdot \bar{\mathbf{J}}:(x) \, ,
\label{eq:ham-SU2-SM}
\end{align}
where $\mathbf{J} = (J^0,J^1,J^2)$ and $\bar{\mathbf{J} }= (\bar{J}^0,\bar{J}^1,\bar{J}^2)$ are the $\mathrm{SU}(2)_1$ Kac-Moody chiral and anti-chiral currents, respectively. Here, $v$ is the velocity, $c=1$ is the central charge, and $g$ is the coupling of the marginal term.

In the free-field representation, the $\mathrm{SU}(2)_1$ Kac-Moody currents are expressed in terms of the compactified boson field $\hat{\varphi} = \phi +\bar{\phi}$ with compactification radius $R=\sqrt{2}$:
\begin{align}
    J^0 (w) &= \frac{i}{\sqrt{2}}\partial\phi (w) \,,\quad J^{\pm} (w) = J^1 (w) \pm i J^2 (w) =:e^{\pm i \sqrt{2}\phi (w)}:\,,\nonumber\\
    \bar{J}^0 (\bar{w}) &= \frac{i}{\sqrt{2}}\bar{\partial}\bar{\phi} (\bar{w}) \,,\quad \bar{J}^{\pm} (\bar{w}) =  \bar{J}^1 (\bar{w}) \pm i \bar{J}^2 (\bar{w}) =:e^{\pm i \sqrt{2}\bar{\phi} (\bar{w})}: \, ,
\end{align}
where $w = e^{\frac{2\pi}{L}(v\tau + i x)}$ is the complex coordinate on the plane.
The unperturbed Hamiltonian $H_0$, corresponding to the $\mathrm{SU}(2)_1$ WZW CFT, represents  a Tomonaga-Luttinger liquid with Luttinger parameter $K=R^2/4=1/2$. The effective coupling $g>0$ in $H_1$ can be understood from the bosonization treatment of the spin-1/2 Heisenberg chain, $S_j^z S_{j+1}^z \sim (\partial_x\hat{\vartheta} (x))^2:  -\alpha^2\left(\frac{2\pi}{L}\right)^2 :\cos [\sqrt{2}\hat{\vartheta} (x)]: $, with $\alpha$ being a non-universal constant. 

Denoting the ground state of effective Hamiltonian [Eq.~\eqref{eq:eff-ham-SM}] as $|\Psi_0(g)\rangle$, its overlap with the conformal crosscap state $|\mathcal{C}_+\rangle$ [Eq.~\eqref{eq:cross-2-SM} with $\kappa_{+} = 2^{1/4}$ for the $\mathrm{SU}(2)_1$ WZW CFT], which has been identified as the continuum counterpart of the lattice crosscap state $|\mathcal{C}_{\mathrm{latt}}\rangle$, should be a universal scaling function of $g$. This overlap, denoted as $|\langle\mathcal{C}_+|\Psi_0(g)\rangle| = F(g)$, can be analytically calculated using the conformal perturbation theory~\cite{Saleur1987,ZhangYS2024}. Using the conformal perturbation formalism developed in Ref.~\cite{ZhangYS2024}, the leading correction to the overlap is given by
\begin{align}
    |\langle \mathcal{C}_+|\Psi_0(g)\rangle| \equiv F(g) =2^{1/4} + \frac{\pi\mathcal{ A}}{2}\cdot g +\mathcal{O}\left(g^2\right)\,,
\label{eq:cross-1st-perturb-SM}
\end{align}
where $\mathcal{A} = -\sqrt{\frac{2}{3}}$ is the amplitude of the one-point crosscap correlator
\begin{align}
    \langle 0,0|\frac{2}{\sqrt{3}} :\mathbf{J} \cdot\bar{\mathbf{J}}:(w,\bar{w})|\mathcal{C}_+\rangle 
    = -\sqrt{\frac{2}{3}} \cdot \frac{1}{(1+|w|^2)^2}\equiv \frac{\mathcal{A}}{(1+|w|^2)^2} \, .
\end{align}
Here $|0,0\rangle$ is the vacuum of the $\mathrm{SU}(2)_1$ WZW CFT.

The perturbation operator $\frac{2 }{\sqrt{3}}:\mathbf{J}\cdot \bar{\mathbf{J}}:$ in Eq.~\eqref{eq:ham-SU2-SM}, with conformal weight $h=\bar{h}=1$, is marginal at first order. The one-loop renormalization group (RG) equation, derived by considering an infinitesimal scale transformation $a \to (1+\delta l)a$ ($a$: short distance cutoff at a given scale) shows that the perturbation is indeed marginal irrelevant~\cite{Cardy1986b,Affleck1989}:
\begin{align}
    \frac{\mathrm{d}g}{\mathrm{d}l} = -\pi b g^2 + \mathcal{O}(g^3) \, ,
\label{eq:RG-SM}
\end{align}
where $b = 4/\sqrt{3}$ is determined by the three-point correlator on the plane
\begin{align}
    \left\langle \frac{2 }{\sqrt{3}}:\mathbf{J}\cdot \bar{\mathbf{J}}:(w_1,\bar{w}_1)\cdot \frac{2 }{\sqrt{3}}:\mathbf{J}\cdot \bar{\mathbf{J}}:(w_2,\bar{w}_2)\cdot \frac{2 }{\sqrt{3}}:\mathbf{J}\cdot \bar{\mathbf{J}}:(w_3,\bar{w}_3)\right\rangle = \frac{-b}{|w_1-w_2|^2|w_1-w_3|^2|w_2-w_3|^2}\,.
\end{align}

To account for the finite-size correction arising from the marginally irrelevant perturbation, we consider a spin-1/2 Heisenberg chain with $N$ sites and lattice spacing $a$ (short-distance cutoff at the ultraviolet scale), with periodic boundary condition imposed. The RG flow is from a theory at the ultraviolet (UV) scale with microscopic coupling $g_{\mathrm{uv}}$ to its low-energy effective description [Eq.~\eqref{eq:eff-ham-SM}] at the infrared (IR) scale with effective coupling $g$. The length of the system $L= Na = N' a'$ is fixed under the RG flow, $e^{l} = \frac{a'}{a} = \frac{N}{N'}$, where $a'$ is the short-distance cutoff at the IR scale. Solving the RG equation [Eq.~\eqref{eq:RG-SM}] gives the effective coupling $g$ as the function of $N$:
\begin{align}
    g\equiv g(l) = \frac{g_{\mathrm{uv}}}{1+\pi b g_{\mathrm{uv}}l}= \frac{g_{\mathrm{uv}}}{1+\pi b g_{\mathrm{uv}}\ln (N/N')}\,.
\end{align}
Specifically, in the limit $\ln N \gg 1/g_{\mathrm{uv}}$, the asymptotic finite-size scaling behavior of the effective coupling $g$ is independent of microscopic details~\cite{Cardy1986b,Affleck1989}:
\begin{align}
    g \to \frac{1}{\pi b \ln N} \, .
\end{align}
This leads to the universal marginal correction of the crosscap overlap [Eq.~\eqref{eq:cross-1st-perturb-SM}]:
\begin{align}
    |\langle \mathcal{C}_+|\Psi_0(g)\rangle| = F(g) \to F(1/\pi b\ln N) = 2^{1/4} - \frac{1}{4\sqrt{2}}\cdot \frac{1}{\ln N} +\mathcal{O}\left(1/(\ln N)^2\right)\,.
\label{eq:cross-scaling-SM}
\end{align}

As discussed above, the universal finite-size correction due to the marginal term is valid for $\ln N\gg 1/g_{\mathrm{uv}}$. However, in lattice simulations, the total number of sites $N$ is typically not sufficiently large to reach that regime. Nevertheless, the form of Eq.~\eqref{eq:cross-1st-perturb-SM} suggests that the lattice crosscap overlap can be expanded in terms of $1/\ln N$, i.e., $|\langle\mathcal{C}_{\mathrm{latt}}|\Psi_0(N)\rangle| = a_1 + a_2/\ln N$, where $|\Psi_0(N)\rangle$ is the ground state of the Heisenberg chain and $a_1$ and $a_2$ are fitting parameters. The fitted Luttinger parameter is given by $K = 1/a_1^4$ (theoretical value for the $\mathrm{SU}(2)_1$ WZW CFT: $K=1/2$).

\end{document}